# The ALMA Science Archive Reaches a Major Milestone

Felix Stoehr[1]
Alisdair Manning[1]
Stewart McLay[1]
Kyoko Ashigatawa[2]
Miguel del Prado[1]
Dustin Jenkins[3]
Adrian Damian[3]
Kuo-Song Wang[4]
Anthony Moraghan[4]
Adele Plunkett[5]
Andrew Lipnicky[5]
Patricio Sanhueza[2]
Gabriela Calistro Rivera[1]
Severin Gaudet[3]

[1] ESO
[2] National Astronomical Observatory of Japan, Mitaka, Japan
[3] Canadian Astronomy Data Centre, Victoria, Canada
[4] Academia Sinica Institute of Astronomy and Astrophysics, Taipei, Taiwan
[5] National Radio Astronomy Observatory, Charlottesville, USA

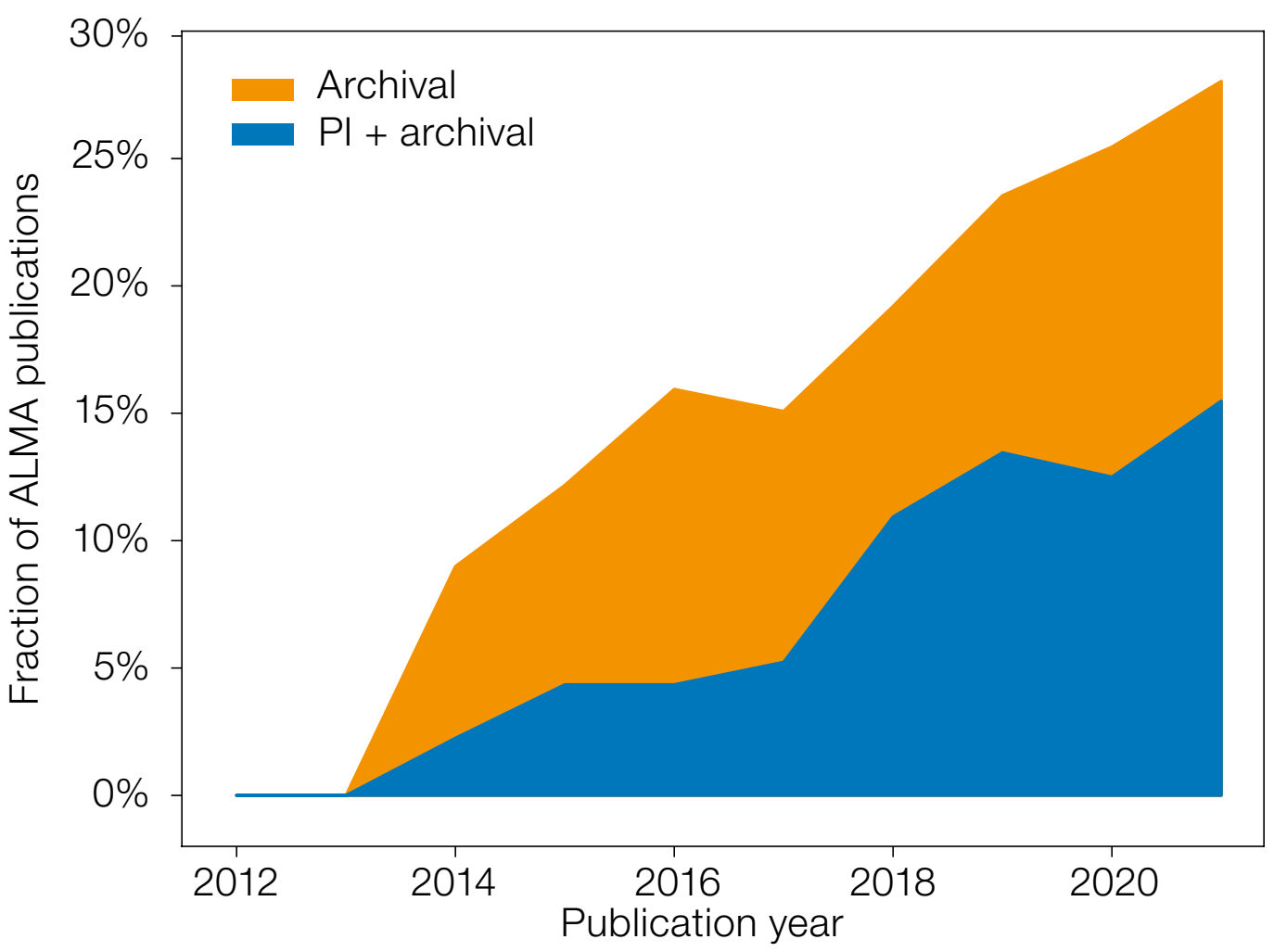

Figure 1. The fraction of ALMA publications with archival context (excluding Science Verification data) is shown as a function of the publication year. The fraction of publications making use of ALMA archival data together with proprietary PI data is shown in blue. Stacked on top is the fraction of publications making exclusive use of ALMA archival data, shown in yellow. Overall, the fraction of ALMA publications with archival context has been continually rising, reaching 28% in 2021.

**Science archives are cornerstones of modern astronomical facilities. In this paper we describe the version 1.0 milestone of the Atacama Large Millimeter/submillimeter Array Science Archive. This version features a comprehensive query interface with rich metadata and visualisation of the spatial and spectral locations of the observations, a complete set of virtual observatory services for programmatic access, text-based similarity search, display and query for types of astronomical objects in SIMBAD and NED, browser-based remote visualisation, interactive previews with tentative line identification and extensive documentation including video and Jupyter Notebook tutorials. The development is regularly evaluated by means of user surveys and is entirely focused on providing the best possible user experience with the goal of helping to maximise the scientific productivity of the observatory.**

## The big picture

Science archives form an integral part of modern observatories. They contribute substantially to the success of an observatory by helping to maximise the scientific output and come with a very favourable cost-benefit ratio; they enable independent scientific research using existing data, and particularly facilitate modern multi-wavelength astronomy as well as studies of time-variable sources. In addition, science archives are the guardians of the fundamental requirement of science to allow for the reproducibility of scientific results, they give access to data for scientists in developing countries (Peek et al., 2019), and they are also used for the proposal preparation process, citizen science, and outreach.

In 2021 28% of all Atacama Large Millimeter/submillimeter Array (ALMA) publications made use of archival data of Principal Investigator (PI) observations, and that fraction is continuously growing (Figure 1). Once high-quality data are generated and processed at an observatory, the main effort in building a valuable and effective science archive is related to data curation, i.e., the extraction, correction, homogenisation, computation and explanation of metadata, as well as the presentation to the users in the form of the reliable search, download, interoperability and analysis functionalities.

The work done on the ALMA Science Archive (ASA) is strictly prioritised according to the overall positive impact anticipated for the users, following ALMA's principle to be easily accessible to all astronomers, very much including those outside the millimetre/submillimetre community. We closely track the wishes of the users both quantitatively (Figure 2) and qualitatively, and have recently finished analysing and discussing, one comment at a time, all 43 pages of comments from the last ALMA user survey dedicated to the ASA. That survey explicitly asked for "wild" and "out-of-the-box" ideas. We also use such ideas, and others, in DesignThinking (DT) workshops, which are a valuable and modern tool for our entirely user-experience-centred design approach.

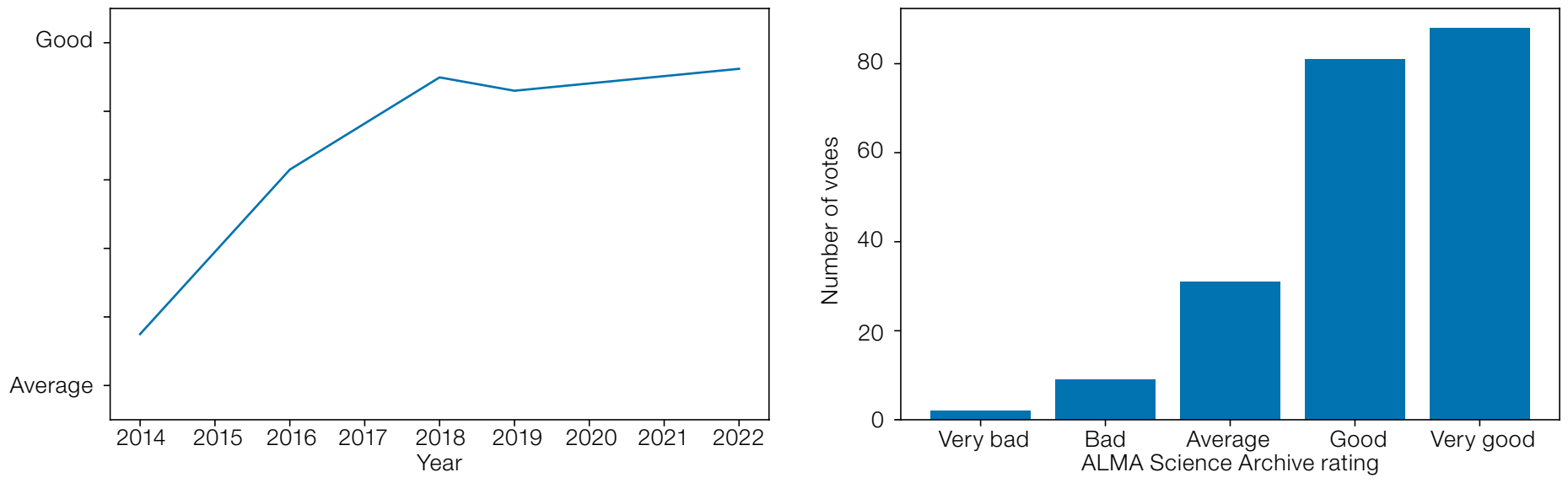

Figure 2. In regular user surveys users are asked to provide comments, but also to rate various aspects of ALMA. While the overall usability of the ALMA Science Archive, rated as "very bad", "bad", "average", "good" or "very good", has progressed significantly over the years, there is still room for improvement (left). The right-hand plot shows the rating distribution for the ASA query interface as of 2019.

One of the surprising general outcomes of those DT workshops was how important it was for users to be able to perform tasks quickly, over just being able to perform the tasks at all or even over being able to perform them easily. We call this need “fastronomy”. We believe that it will gain more importance in the next decade within astronomy and we make it a pillar of our own design strategy.

Allowing users to perform their tasks rapidly is even more challenging as the astronomical data taken grow exponentially with time (Stoehr, 2019). Therefore, a continued effort has to be made to keep up the pace by ensuring that all tools scale accordingly, and in particular to solve what is the real Big Data problem: reducing the fraction of data that astronomers have to actually look at. This effort will inevitably come with increased responsibility on the part of the observatories and will probably have to rely substantially on artificial intelligence techniques in the long-term future (Stoehr, 2019).

Currently the ASA query interface is accessed from 6000 distinct IP addresses each quarter. The ASA holds data from about 53 000 science observations distributed over 49 million files, totalling 1.3 PB in size. About 50 to 100 TB of data are served to the users every month with a healthy ratio of data downloaded by users versus data ingested from the telescope; this ratio is currently about a factor of three (Figure 3, left). The vast majority of the data are already out of their 12-month proprietary period when they are downloaded (Figure 3, right). So far, ALMA data have been used in over 2700 publications.

In the remainder of this paper, we describe our recent work which led to the milestone of the ASA we call version 1.0. Please refer to some of our previous work for astronomical archives in general (Stoehr, 2019), for the ALMA Science Archive (Stoehr et al., 2017) or for the principles of user interface design (Stoehr, 2017).

## Features

The ASA query interface[1] offers search-as-you-type and instantaneous results from the entire multidimensional metadata cube, i.e., unscoped access to all 49 columns. That metadata cube contains extensive information about the observations, but also information about the corresponding projects and about the ALMA publications, for example authors or abstracts. The latter is possible because the library and information services at ESO, the National Radio Astronomy Observatory (NRAO) and the National Astronomical Observatory of Japan (NAOJ) carefully track each publication and record the data that have been used (for example, Grothkopf, Meakins & Bordelon, 2018), which authors are required by policy to specify in the publication’s acknowledgment section. Queries can combine search constraints on any of those three metadata categories. We offer users the possibility of uploading a list of targets or coordinates, showing the number of times a particular dataset was used in a publication, allowing modification of the layout of the interface to fit the users’ needs, enabling column reordering as well as sorting the entire data holdings and providing a bookmark to the current search and settings that users can save or share with colleagues. In all our designs the goal is always to show only the most relevant information while at the same time allowing experts to drill down to the metadata level they need.

Substantial effort has been put into combining the values of the raw data into one row per observed source, and allowing for searches on highly relevant scientific properties, i.e., real physical parameters of the observations, such as the estimated continuum and line sensitivity (using the ALMA sensitivity calculator with parameters of the observed raw data), the integration time, the frequency resolution, or the expected spatial resolution and maximum recoverable spatial scale. To achieve homogeneity and consistency of the metadata, at each major software release all metadata are fully recomputed from the original raw data.

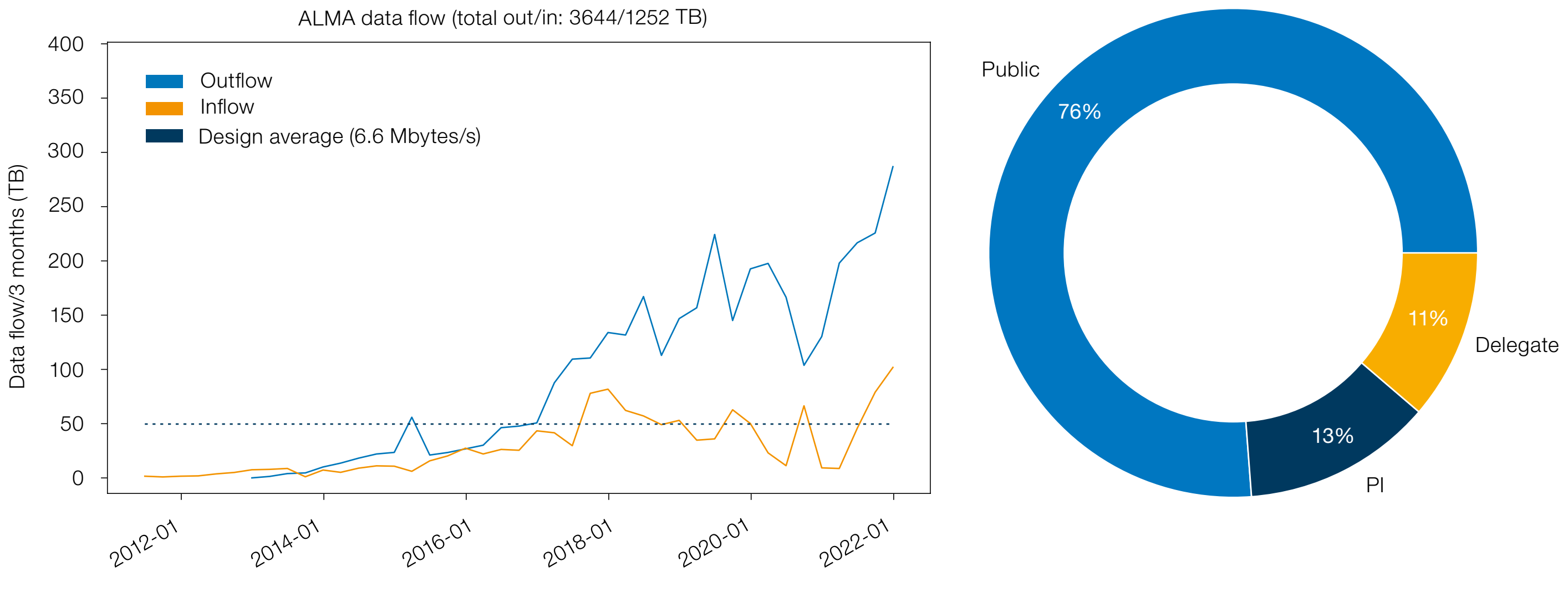

Figure 3. The left panel shows the data inflow (observations and data products) and data outflow (data download) per quarter. The dashed line indicates ALMA’s current design average data rate. The right panel shows which fraction of the data were downloaded when they were public compared to the fractions of the downloads made during the proprietary period by the PI as well as by a person to whom the PI has delegated access to the data.

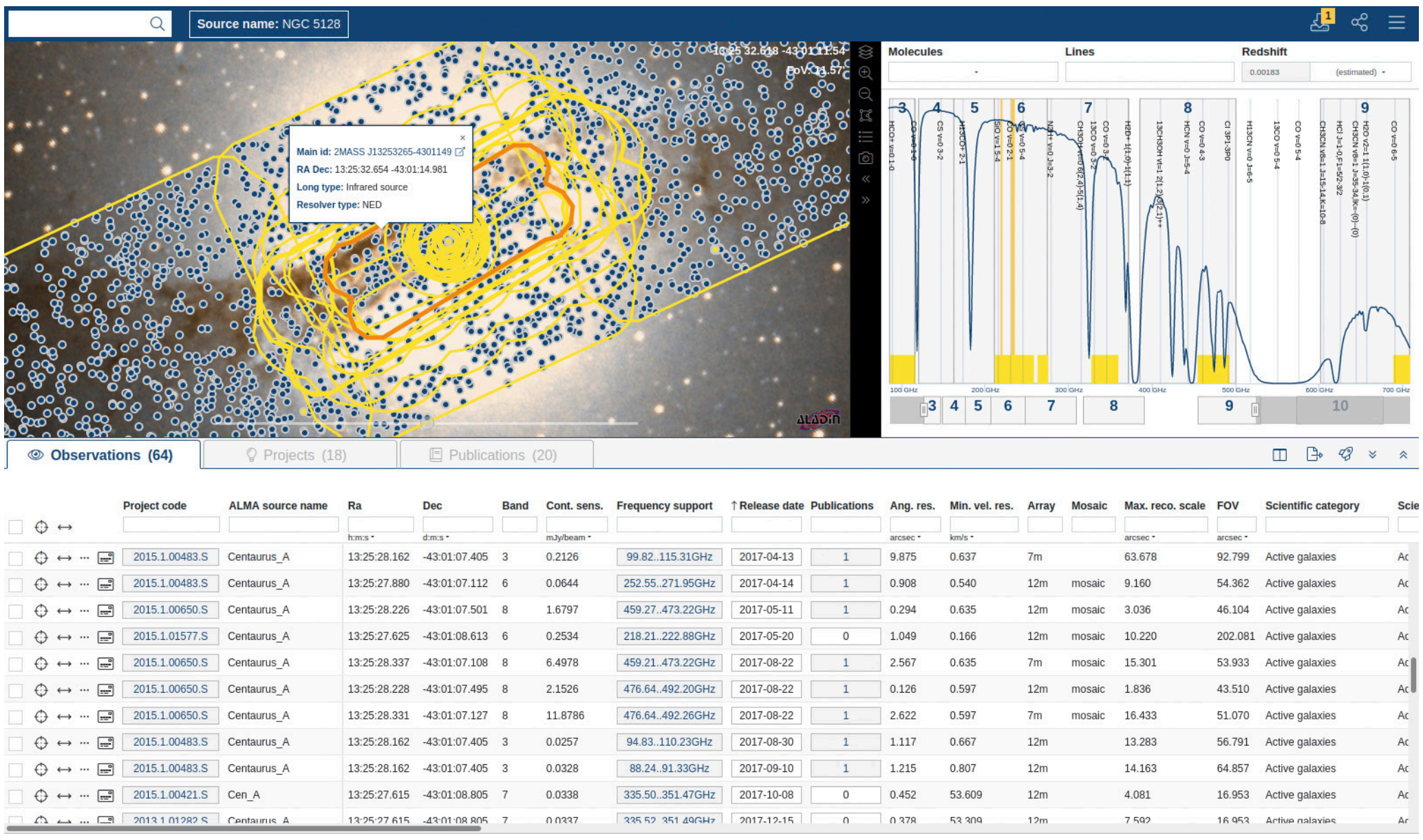

Figure 4. The ASA query interface features search-as-you-type, a sky view, a spectral coverage viewer, a result table, and much more. All objects from SIMBAD and NED falling into any of the regions of the ALMA observations are displayed and even the object-types can be queried for.

In addition to AladinLite[2], the query interface makes extensive use of modern web technologies, in particular the Angular web[3] framework as well as ElasticSearch[4]. We have implemented virtual scrolling on top of those technologies to offer a natural user experience and full scalability. Help is provided throughout the interface by means of tooltips, video tutorials[5] and the extensive Science Archive Manual[6].

## Text similarity search

Users are often interested in projects or publications similar to the one they are looking at, but discovering those used not to be straightforward. To this end, the ASA has now implemented a state-of-the-art text-similarity-based recommender system (suggestion and proof-of-concept: Alejandro Barrientos) on all projects and publications. The “you might also be interested in” lists can be placed into a new browser tab allowing the users to add further search constraints.

## Astronomical context

Many ALMA observations are not just 2D images but are full 3D datacubes, which are challenging to present on the interface. We have developed a spectral coverage viewer to show the exact extent of the frequency coverage of each observation, and even the extent for entire projects and/or publications. From the values PIs enter into the ALMA Observing Tool (OT) when preparing their proposals, we estimate the median redshift of the displayed results and then show the most relevant velocity-shifted spectral line transitions with the possibility of seeing more transitions by zooming-in further, or limiting the drawn lines to predefined categories (for example, Hot Cores) or species.

The spatial location and extent of all ALMA observations are displayed in the Aladin Sky view. In addition, observations can be directly selected from that view. The result table automatically scrolls to the corresponding position. Also through the sky view we provide additional astronomical context; we have implemented a slider that allows users to fade smoothly through the entire frequency spectrum where selected background data from sources such as the Digitized Sky Survey, the Spectral and Photometric Imaging REceiver (SPIRE) on the Herschel Space Observatory or the Atacama Cosmology Telescope/Planck maps are downloaded from the Strasbourg astronomical Data Center (CDS7) in HiPS format and displayed on the fly. Users can also select additional backgrounds from the layers icon.

At high zoom levels, the ASA sky view also displays all SIMBAD[8] and NED[9] sources that fall into any of the regions observed by ALMA. Tooltips show information about the sources and provide links to NED and SIMBAD for further details.

## Object type search

The ability to query by object type (for example, “Active Galaxy Nucleus” or “Galaxy pair”) rather than a specific target, together with any other search constraint, is challenging to implement but is motivated by an important use case, demonstrated by user feedback. A new feature enables this type of object-type search now that the metadata from SIMBAD and NED have been retrieved and fully integrated into the ALMA database (Figure 4).

By default all regions are returned where a source of that particular type is in the field of view, regardless of whether or not ALMA may have actually detected it. In addition to that default setting, we have tried to identify the “best match” object out of the many SIMBAD or NED objects falling into each of the ALMA observation footprints by taking into account the observed emission strength, source name and position on the sky given by the PI. Users can thus select object types and restrict the search to those that have likely been the main targets of the observations. In the sky view display, these best matching objects are marked in yellow.

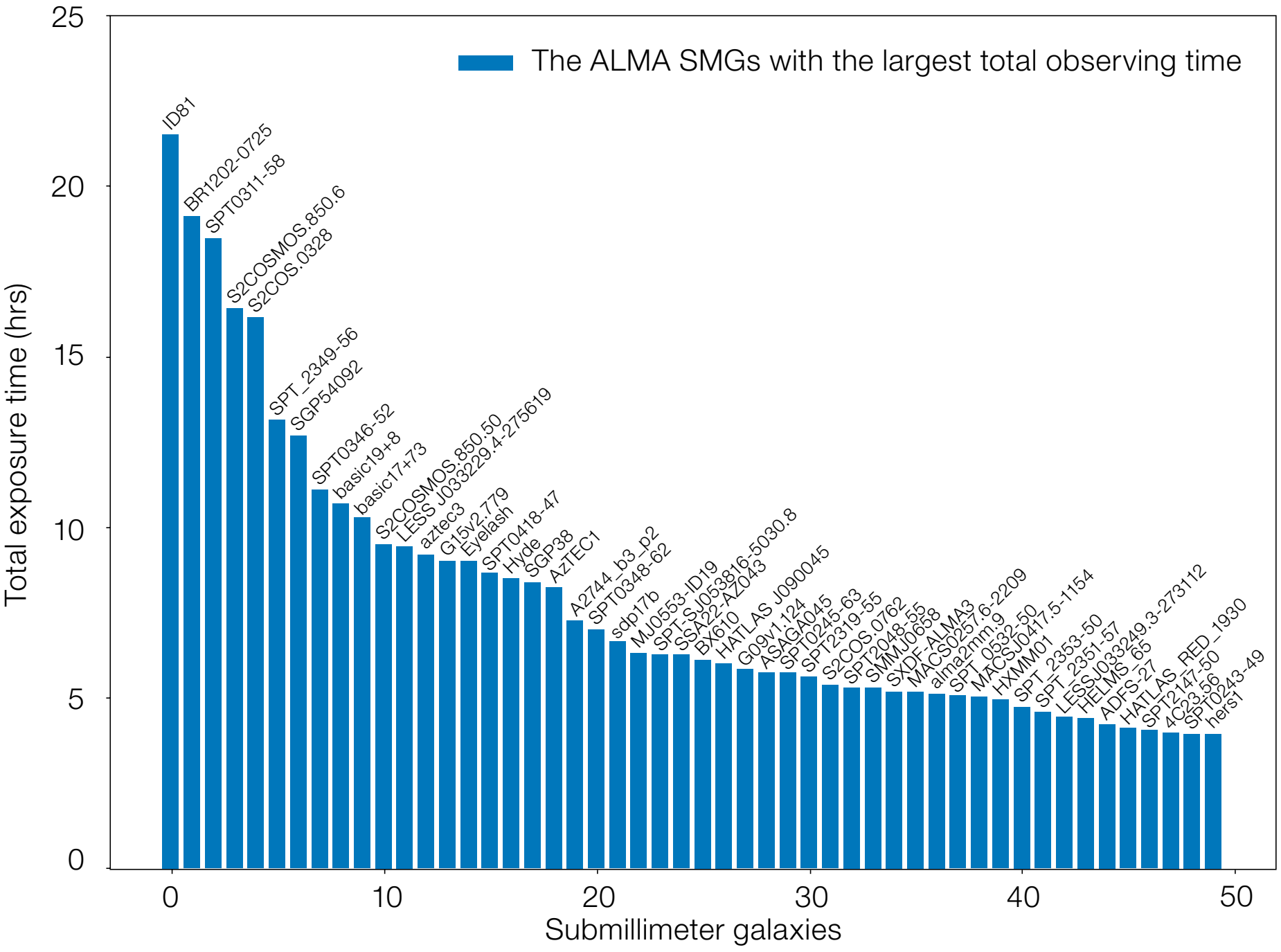

Figure 5. The ASA can be accessed via a web interface but also completely programmatically, i.e., through VO protocols and standards. We provide Jupyter Notebooks which show, for example, that with a few lines of Python this plot can be created, identifying submillimeter galaxies (SMGs) that have been observed most with ALMA, adding up the observing time across all observing cycles.

## Virtual observatory

The ASA is now fully interoperable through virtual observatory (VO) standards. Using the rocket icon on the query interface, the query results can be broadcast through the VO SAMP protocol to other VO tools like Aladin[10] and Topcat[11]. The results can also be exported directly as VO Tables.

Moreover, the ASA provides a simple image access through the SIAv2 protocol[12], ObsCore[13] access through the TAP protocol[14], data exploration and download through the DataLink protocol[15], and also — our most recent addition — cutouts of FITS cubes and images via the SODA protocol[16]. These standardised interfaces can all be discovered through the registry services of the IVOA[17] and allow users to access the ASA in a fully programmatic way, including for data mining and machine learning, with the tools of their choice. We provide extensive Jupyter Notebook tutorials[18] demonstrating access to the ASA through the VO protocols using Python (Figure 5).

## Previews

Previews — quick-look images allowing users to grasp the content of the data files at a glance — are a highly desired ingredient of modern astronomical archives. A particular challenge for ALMA is to provide previews that give the maximally useful essence of large 3D data cubes with up to 3800 spectral channels. After substantial experimenting, we opted for previews as shown in Figure 6. The previews are static and fully self-contained interactive html files. Users can zoom and pan the images and spectra for which we make use of the bokeh.org[19] library. Data products can be separated into two parts, the continuum emission (the flux for a pixel that is constant throughout the entire data cube) and the line emission (the remainder after the continuum was subtracted). Just collapsing (averaging) the line emission to create a two-dimensional image often erases the structure of the emission that sometimes can be concentrated into only a few spectral channels. We find that the peak-flux image (for each pixel the largest value along the spectral axis, (sometimes also called "moment 8") gives the most useful representation of the line emission.

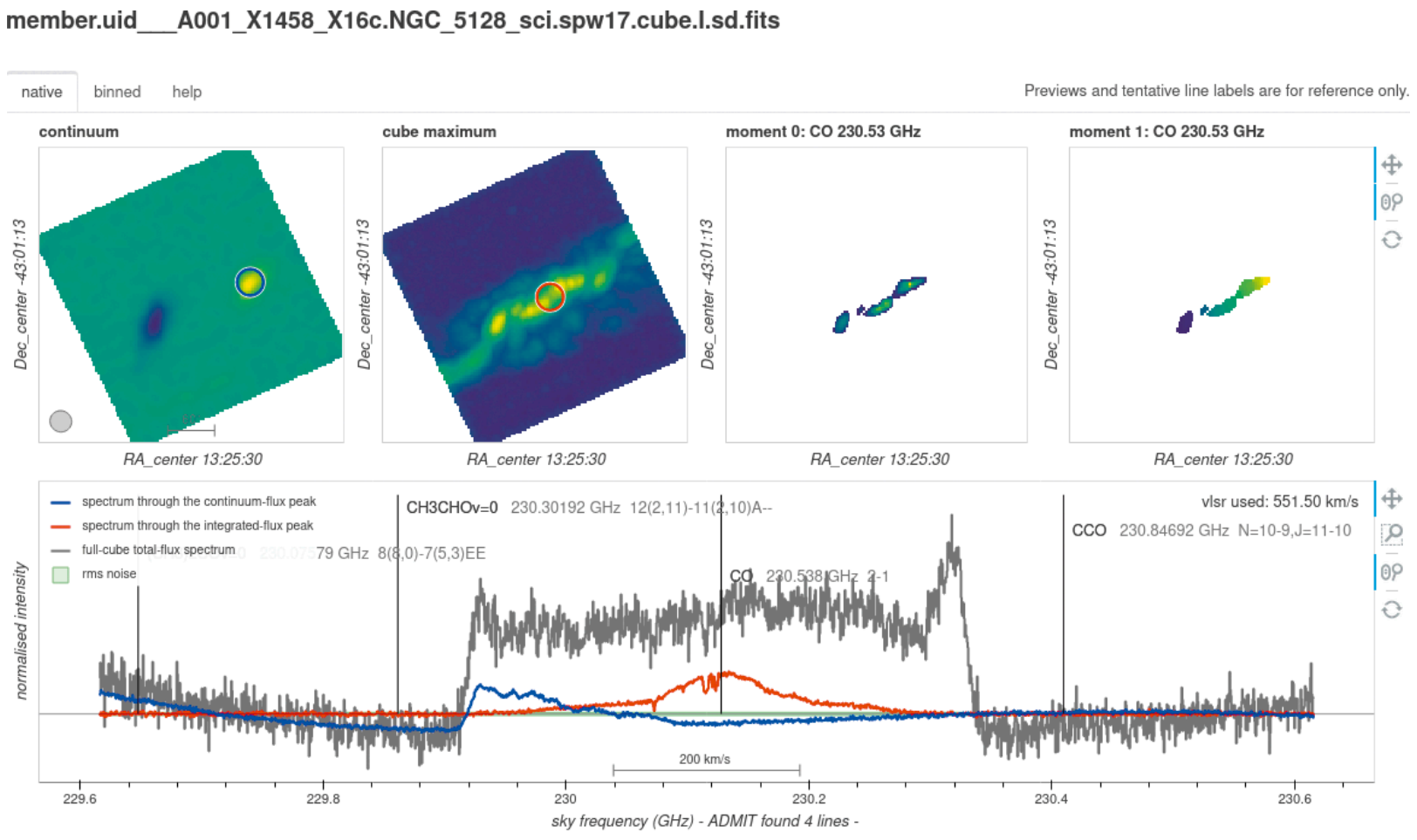

Figure 6. The ALMA previews show the continuum emission, the peak flux image (also called moment-8), and the moment-0 and moment-1 maps of the strongest detected line. In the lower part, three cuts through the datacube are presented. We use the ALMA Data Mining Toolkit (ADMIT) to identify spectral features and to tentatively label them. Those features are overplotted as vertical lines. Tabs provide access to the same analysis but binned by a factor of 16 in the spectral dimension, as well as to an extensive help. All panels are interactive and allow for zooming and panning.

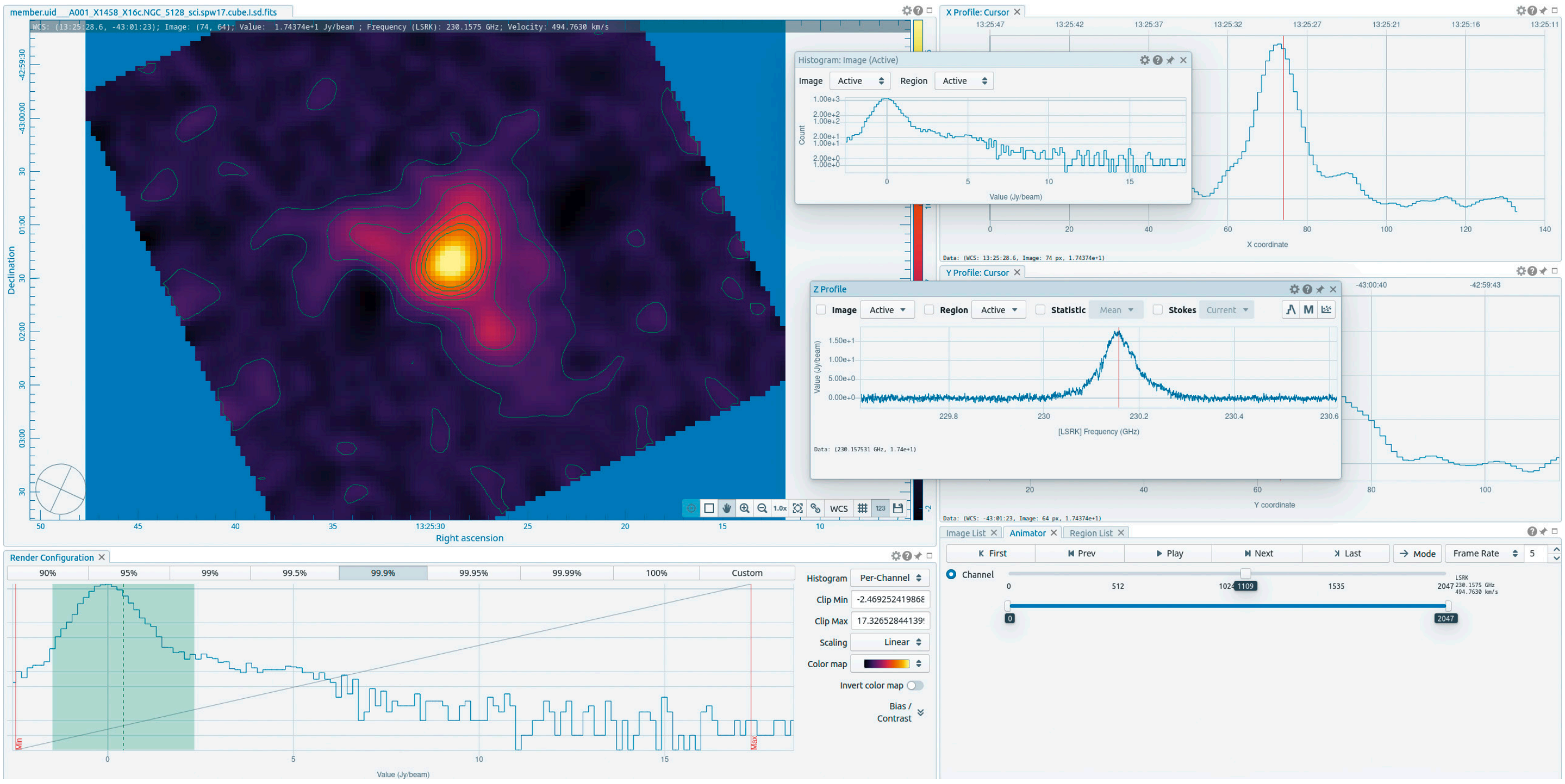

Figure 7. ALMA has implemented 1-click remote visualisation of all FITS files in the ASA using the feature-rich and sophisticated CARTA software. Powerful servers at each of the ALMA ARCs serve the minimally necessary data to the user's web browser providing a real-time user experience, even for extremely large data cubes.

In addition to showing those images, we run the ALMA Data Mining Toolkit (ADMIT)[20] over the data cubes. That software, originating in an ALMA development project, tries to find spectral lines in the data cube and — if provided with a guess for the relative velocity or redshift of the observed source — to identify them and label them with the molecule and transition tentatively responsible for the line emission. We use the following algorithm to provide the velocity to the ADMIT software. If a velocity was given by the PI in the ALMA OT, that velocity is used. If no velocity was given, we look for other existing ALMA observations of the same source and use the velocity if a value is available. If not, we try to match the observed source with a source from SIMBAD and NED and use the velocity from those services. If none is available, we check to see if the PI has entered the lines they were looking for in the OT and compute the velocities from that information. This process provides guesses of velocities for about 80% of all FITS files.

Three spectra across the cube are shown in the spectral panel of the preview. In the case that strong lines have been detected, the moment-0 and moment-1 images of the strongest line are displayed. The same plots do also exist in a binned version on the preview which makes the detection of weak broad lines easier. Finally, a help tab provides detailed descriptions of the content.

While not all lines can be detected and the line labels are tentative, using ADMIT across all FITS files is the first step towards automated data-annotation of ALMA data, i.e., the first step towards what is expected to be the next frontier for non-survey-type astronomical archives (Durand, private communication): describing the actual content of the data rather than only the observations themselves. Automatically described content has the potential to hugely speed up the data-exploration process and is thus fully inline with the "fastronomy" concept.

## CARTA

The Cube Analysis and Rendering Tool for Astronomy (CARTA)[21] is an extremely powerful and feature-rich science-grade visualisation tool with a fast-paced and well-funded development process (Figure 7). The same software can be installed as a standalone desktop application or run partly on a server at a facility like ALMA and partly on the user's machine, creating a very fast user experience. Users can then connect to the server remotely and visualise archival images or cubes directly in their web browser without having to download the data at all. But they can still experience a fully seamless user experience identical to their CARTA desktop installation. In the client server setup, the server is located very close to the data and performs all the data access and pre-processing such that only the absolute minimum information has to be sent over the internet to the users. The result is that FITS cubes, even those above 1TB in size, can be opened in seconds, and visualised and analysed in real-time. Next to the spectral coverage viewer and the previews, CARTA is our third pillar helping with the 3D data challenge. In collaboration with the CARTA team, ALMA has deployed one server instance at each of the three ALMA Regional Centres (ARCs) on dedicated hardware (64 cores, 512 GB RAM, as well as a powerful GPU). One-click visualisation is available for users directly from the preview window of the query interface or from the Request Handler interface, again part of "fastronomy".

## Request Handler

Once data of interest have been selected, they can be downloaded. Unless data are still protected by the proprietary period, typically 12 months, all interaction and download can happen anonymously. The Request Handler window allows users to see the products for the selected data showing the minimum matching number of products per default. With tabs, users can decide to view larger parts of the product hierarchy. The products can also be selected in categories or filtered by name. All files can be downloaded individually or conveniently packaged into .tar files.

As ALMA data can be large, we offer download through a script that carries out the download in five parallel streams. This script can also be executed elsewhere, for example on a processing cluster. The user can decide whether or not the data should be unpacked and, if so, whether or not the directory structure of the dataset shall be preserved.

The Request Handler provides products created by ALMA as well as externally contributed products, like the products from the Additional Representative Images for Legacy (ARI-L) project (Massardi et al., 2021) and products provided by the PIs of Large Programmes.

## Outlook

The development of the ASA is of course not stopping at the version 1.0 milestone. On the contrary, in addition to the constant adaptation of the ASA to changes in ALMA's capabilities and policies, a number of new features are in the near-term development plan, the largest one being the integration of processing of all ALMA FITS products with ADMIT into the ALMA data workflow at the Joint ALMA Observatory (JAO). As a result, the metadata of detected and then tentatively labelled lines can be made available for queries.

Additional resources would enable the implementation of ambitious ideas for the long-term evolution of the ALMA data flow and of the ASA, including: high-level data products; user-initiated remote creation of calibrated measurement sets; user-initiated remote imaging and data combination; regular automated bulk reprocessing of the data holdings with the latest version of the ALMA pipeline; and a fully fledged ALMA Science Platform including access to graphics processing units for machine learning.

### Acknowledgements

Over the past 12 years, a number of current and former colleagues have been part of the computing and science teams of the ASA and their work is very gratefully acknowledged: Christophe Moins, Matthias Bauhofer, Mark Lacy, Stéphane Leon Tanne, Brenda Matthews, Erik Muller, Masao Saito, Eric Murphy, Juande Santander Vela, John Hibbard, Akiko Kawamura, Tsuyoshi Kobayashi and Dilip Diascore. We thank the ADMIT team for all their help. We are also indebted to the library staff at ESO, NRAO and NAOJ for tracking all the publications, to the archive operation staff at the JAO and at the ARCs for running all the systems as well as to all ALMA staff, including at the ARC nodes, for their continued work and support. ALMA is a partnership of ESO (representing its Member States), NSF (USA) and NINS (Japan), together with NRC (Canada), MOST and ASIAA (Taiwan), and KASI (Republic of Korea), in cooperation with the Republic of Chile. The JAO is operated by ESO, AUI/NRAO and the NAOJ.

### References

Grothkopf, U., Meakins, S. & Bordelon, D. 2018, EPJ Web of Conferences, 186, 06001
Massardi, M. et al. 2021, PASP, 133, 1026
Peek, J. et al. 2019, BAAS, 51, 105
Stoehr, F. 2017, in ASP Conf. Ser., 512, Astronomical Data Analysis Software and Systems XXV, ed. Lorente, N. P. F., Shortridge, K. & Wayth, R., (San Francisco: ASP), 511
Stoehr, F. 2019, in ASP Conf. Ser., 523, Astronomical Data Analysis Software and Systems XXVIII, ed. Teuben, P. J., Pound, M. W., Thomas, B. A. & Warner, E. M., (San Francisco: ASP), 378
Stoehr, F. et al. 2017, The Messenger, 167, 2

### Links

[1] ASA web interface: https://almascience.org/aq
[2] Angular web framework: https://angular.io
[3] AladinLite: https://aladin.u-strasbg.fr/AladinLite
[4] Elastic search: https://www.elastic.co
[5] ASA video tutorials: https://almascience.org/alma-data/archive/archive-video-tutorials
[6] ASA manual: https://almascience.org/documents-and-tools/latest/science-archive-manual
[7] CDS: https://cds.u-strasbg.fr
[8] SIMBAD: http://simbad.u-strasbg.fr/simbad
[9] NED: https://ned.ipac.caltech.edu
[10] Aladin: https://aladin.u-strasbg.fr
[11] Topcat: http://www.star.bris.ac.uk/~mbt/topcat
[12] SIAv2: https://www.ivoa.net/documents/SIA/20151223
[13] ObsCore: https://www.ivoa.net/documents/ObsCore/20170509/index.html
[14] TAP: https://www.ivoa.net/documents/TAP/20190927
[15] DataLink: https://www.ivoa.net/documents/DataLink
[16] SODA: https://www.ivoa.net/documents/SODA/20170517/index.html
[17] EuroVO Registry: http://registry.euro-vo.org/evor
[18] ASA Jupyter Tutorials: https://almascience.eso.org/alma-data/archive/archive-notebooks
[19] Bokeh library: https://bokeh.org
[20] ADMIT: https://admit.astro.umd.edu
[21] CARTA: https://cartavis.org